# Regression to the Tail: Why the Olympics Blow Up

By Bent Flyvbjerg,[a] Alexander Budzier,[b] and Daniel Lunn[c]

a) BT Professor and Chair of Major Programme Management, University of Oxford's Saïd Business School, corresponding author.
b) Fellow in Management Practice, University of Oxford's Saïd Business School.
c) Fellow at Worcester College, Oxford, and Professor of Statistics Emeritus at University of Oxford's Department of Statistics.

**Abstract**

The Olympic Games are the largest, highest-profile, and most expensive megaevent hosted by cities and nations. Average sports-related costs of hosting are $12.0 billion. Non-sports-related costs are typically several times that. Every Olympics since 1960 has run over budget, at an average of 172 percent in real terms, the highest overrun on record for any type of megaproject. The paper tests theoretical statistical distributions against empirical data for the costs of the Games, in order to explain the cost risks faced by host cities and nations. It is documented, for the first time, that cost and cost overrun for the Games follow a power-law distribution. Olympic costs are subject to infinite mean and variance, with dire consequences for predictability and planning. We name this phenomenon "regression to the tail": it is only a matter of time until a new extreme event occurs, with an overrun larger than the largest so far, and thus more disruptive and less plannable. The generative mechanism for the Olympic power law is identified as strong convexity prompted by six causal drivers: irreversibility, fixed deadlines, the Blank Check Syndrome, tight coupling, long planning horizons, and an Eternal Beginner Syndrome. The power law explains why the Games are so difficult to plan and manage successfully, and why cities and nations should think twice before bidding to host. Based on the power law, two heuristics are identified for better decision making on hosting. Finally, the paper develops measures for good practice in planning and managing the Games, including how to mitigate the extreme risks of the Olympic power law.

**Why Olympic Costs Are Important**

The Olympic Games are the largest, the highest-profile, and the most expensive megaevent that exists. The five most recent Games for which data are available, held over the decade 2007-2016, cost on average $12 billion – not including road, rail, airport, hotel, and other infrastructure, which often cost several times that.[1] The financial size and risks of the Games therefore warrant study. As part of bidding for hosting the Games, the International Olympic Committee (IOC) requires host cities and governments to guarantee that they will cover possible overruns to the giant Olympic budgets. This means that hosts get locked in to a non-negotiable commitment to cover such increases. If overruns are likely, as our data show they are, then financial risks should be taken explicitly into account when cities decide whether to bid to host the Olympic games.

Furthermore, given a global economic climate of austerity in government spending, likely to grow even more frugal after the splurge on the covid-19 pandemic, understanding the financial and economic implications of major investments like the Games is critical for governments to make sound decisions about their expenditures. For instance, cost overrun and associated debt from the Athens 2004 Games weakened the Greek economy and contributed to the country's deep financial and economic crises, which began in 2007 and was dubbed the "forever crisis" by *The Financial Times*, because it seemed to never end (Flyvbjerg 2011, Kuper 2017). Similarly, in June 2016 – less than two months before the Rio 2016 opening ceremony – Rio de Janeiro's governor declared a state of emergency to secure additional funding for the Games. When Rio decided to bid for the Olympics, the Brazilian economy was doing well. Now, almost a decade later, costs were escalating and the country was in its worst economic crisis since the 1930's with negative growth and a lack of funds to cover costs. This is a common pattern due to the cyclical nature of recessions: if the economy had high GDP growth when a city decided to bid for the Games, which is typical, then during the seven to 11 years until the opening ceremony the economy is likely to weaken, as shown in Figure 1. Potential hosts – and especially those with small and fragile economies – will want to make sure they do not end up like Athens and Rio. They can protect themselves by undertaking a realistic assessment of costs and cost risks before they bid for the Games. The results presented here will allow such assessment in a rigorous, evidence-based manner.

[Figure 1 app. here]

Unfortunately, Olympic officials and hosts often misinform about the costs and cost overruns of the Games. For instance, in 2005 London secured the bid for the 2012 Summer Games with a cost estimate that two years later proved inadequate and was revised upwards with around 100 percent. Then, when it turned out that the final outturn costs were slightly below the revised budget, the organizers falsely, but very publicly, claimed that the London Games had come in under budget, and media uncritically reported this, including the BBC (2013). Such deliberate misinformation of the public about cost and cost overrun treads a fine line between spin and outright lying. It is unethical, no doubt, but very common. We therefore cannot count on organizers, the IOC, and governments to provide us with reliable information about the real costs, cost overruns, and cost risks of the Olympic Games. Independent research is needed, like that offered here.

Previous academic research, including our own, has established that hosting the Olympics is highly costly and financially risky. There is a gap in the literature, however, regarding the specific nature of the risks faced by cities and nations in deciding whether to host the Games or not, and the implications of those risks. With the present paper we aim to fill this gap. First, we present our data for Olympic costs and cost overruns since 1960. Second, we fit different theoretical statistical distributions to the data. Finally, we interpret the best fits in terms of what they say about cost risk and how such risk may be mitigated and managed. Annex 1 contains more detail about previous academic research on cost and cost overrun at the Games and how this relates to the present research.

**Olympic Costs 1960-2016**

We measure Olympic costs as actual outturn sports-related costs of hosting the Games. Non-sports-related, wider capital costs are not included, although they are typically substantial. Cost overruns are measured as actual outturn cost in percent of estimated costs. For example, if a Games was estimated to cost $8 billion and actually cost $12 billion, then these Games incurred a cost overrun of 50 percent. The baseline for measuring cost overrun is the cost estimate at the time of bidding for the Games. All costs are measured in real terms, i.e., not including inflation. All Games from 1960 to 2016, for which data were available, are included, which is 25 out of 30 Games for outturn cost and 19 out of 30 Games for cost overrun. It is an interesting finding in its own right that for 11 out of 30 Games valid and reliable data on cost overrun could not be found, despite trying hard. Incredible as it may sound, for more than a third of Games since 1960 no one seems to know what estimated and actual costs were. Nevertheless, the dataset is the largest and most consistent of its kind, allowing statistical analysis. Annex 2 contains more detail about data and methodology.

Table 1 shows actual outturn sports-related costs of the Olympic Games 1960-2016 together with the number of events and number of athletes in each Games.[2] [3]

[Table 1 app. here]

We see that the most costly Summer Games to date are London 2012 at $15.0 billion. For the Winter Games, Sochi 2014 is the most costly at $21.9 billion. The least costly Summer Games are Tokyo 1964 at $282 million; the least costly Winter Games, Innsbruck 1964 at $22 million.

Average cost for Summer Games 1960-2016 is $6.0 billion (median $5.6 billion). Average cost for Winter Games over the same period is $3.1 billion (median $2.0 billion). The large difference between average and median cost for the Winter Games is mainly caused by the extreme cost of Sochi 2014, which at $21.9 billion cost more than all previous Winter Games combined. Indeed, the Sochi 2014 Winter Olympics are the most costly Games ever, even when compared with the Summer Games. This is extraordinary, given that cost for the Winter Games is typically much lower than for the Summer Games, with the median cost for Winter Games being less than half the median cost for Summer Games. We will return to the importance of extreme values below as it turns out these are typical of the Games to a degree where they define the particular cost risk profile of the Games.[4]

Average cost for all Games 1960-2016 is $4.5 billion (median $2.5 billion). Average cost for the five Games held in the decade 2007-2016 is $12.0 billion (median $13.7 billion). It should again be remembered that wider capital costs (OCOG indirect costs, see Annex 2) for urban, transportation, and business infrastructure are not included in these numbers and that such costs are typically substantial.

Figure 2 shows the development of cost 1960-2016. The trend lines indicate that the cost of the Games have increased over time. Conventional statistical analysis shows that the apparent increase is statistically overwhelmingly significant ($p < 0.001$, Winter Olympics; $p < 0.001$, Summer Olympics).[5] However, due to the existence of extreme values in the data, documented below, results of conventional statistical analysis must be interpreted with caution.

[Figure 2 app. here]

Table 2 shows cost per event and cost per athlete 1960-2016 in 2015 USD. These data were available for 25 of the 30 Games 1960-2016. The average cost per event for the Summer Games is $22.4 million (median $19.7 million). For the Winter Games it is $39.2 million (median $29.5 million). – The highest cost per event in the Summer Games was found for London 2012 at $49.5 million. For the Winter Games, the highest cost per event was found for Sochi at $223.4 million, which again is an extreme value. The lowest cost per event was found for Tokyo 1964 at $1.7 million for the Summer Games and Innsbruck 1964 at $0.6 million for the Winter Games.

[Table 2 app. here]

For cost per athlete, we found the Winter Games to be twice as costly as the Summer Games. The average cost per athlete is $671,000 for the Summer Games (median $547,000) and $1.3 million for the Winter Games (median $990,000). The highest cost per athlete in the Summer Games was found for London 2012 at $1.4 million. For the Winter Games, the highest cost per athlete was found for Sochi 2014 at $7.9 million. The lowest cost per athlete in the Summer Games was found for Tokyo 1964 at $55,000, and in the Winter Games for Innsbruck 1964 at $20,000.

Figure 3 shows the correlation of cost per athlete with time. We see a shift in trend from cost per athlete being generally higher for the Summer than for the Winter Games until the mid 1980's, after which the Winter Games become more costly than the Summer Games, in terms of cost per athlete. We also see that cost per athlete was generally decreasing for the Summer Games from the mid-1980's until the early noughties, after which cost per athlete has been increasing for both the Summer and Winter Games, driven mainly by London 2012 and Sochi 2014.

[Figure 3 app. here]

**Cost Overrun at the Games**

Table 3 shows percentage cost overrun in real terms (not including inflation) for the Olympic Games 1960-2016. Data on cost overrun were available for 19 of the 30 Games 1960-2016. Statistical tests of the difference between bid budgets and final costs show this difference to be statistically overwhelmingly significant ($V = 190$, $p < 0.0001$). That is to say, cost overruns are statistically overwhelmingly manifest for the Olympics. It should be mentioned that if the cost overruns had been calculated in nominal terms (including inflation) they would have been significantly larger. In this sense the numbers shown are conservative.

[Table 3 app. here]

We find the following averages and medians for cost overrun in real terms:

- Summer Games: average cost overrun is 213 percent (median 120 percent).
- Winter Games: average cost overrun is 142 percent (median 118 percent).
- All Games: average cost overrun is 172 percent (median 118 percent).

Even though the difference between average cost overrun for the Summer and Winter Games is relatively large at 71 percentage points, the difference is statistically non-significant (non-parametric test, W = 48, p = 0.778). In statistical terms there is therefore no difference between cost overrun in the Summer and Winter Games and the data may be pooled for statistical analyses, for instance in making more accurate reference class forecasts of budgets for future Olympic Games (Flyvbjerg 2008).

We further see that:

- 15 of 19 Games (79 percent) have cost overruns above 50 percent.
- 10 of 19 Games (53 percent) have cost overruns above 100 percent.

Judging from these statistics it is clear that large risks of large cost overruns are inherent to the Olympic Games. In the next section we will explore further what type of risk is at play and what it entails for the decision to host the Games.

For the Summer Games, the largest cost overrun was found for Montreal 1976 at 720 percent. The smallest cost overrun for the Summer Games was found for Beijing 2008 at two percent. For the Winter Games, the largest cost overrun was for Lake Placid 1980 at 324 percent and the smallest for Vancouver 2010 at 13 percent.

The vigilant reader may be skeptical that the lowest cost overrun of all Games would be for Beijing 2008 at two percent. China is known for its lack of reliability in economic reporting (Koch-Weser 2013). However, the total sports-related cost of $6.8 billion and the cost per athlete of $622,000 for the Beijing 2008 Games are higher than for the majority of other Summer Games (see Tables 1 and 2). The reported costs are therefore deemed adequate for hosting the Beijing Games and we have no evidence that the official numbers have been manipulated and should be rejected. Like other observers of economic data from China we therefore include the numbers, with the caveat that they are possibly less reliable than those from other nations, given the lack of transparency in Chinese economic data. Again, this means that our averages for cost overrun in the Games are likely to be conservative.

In sum, the financial and economic risk of hosting the Olympic Games holds policy and political consequences because:

1. *All Games, without exception, have cost overrun.* For no other type of megaproject is this the case. For other capital investment types, typically 10-20 percent of investments come in on or under budget. For the Olympics it is zero percent. It is worth considering this point carefully. A budget is typically established as the maximum – or, alternatively, the expected – value to be spent on a capital investment. However, in the Games the budget is more like a fictitious minimum that is consistently overspent. Further, even more than in other megaprojects, each budget is established with a legal requirement for the host city and government to guarantee that they will cover the cost overruns of the Games. Our data suggest that this guarantee is akin to writing a blank check for the event, with certainty that the cost will be more than what has been quoted. We call this the "Blank Check Syndrome." In practice, the bid budget is more of a down payment than it is a budget; further installments will follow, written on the blank check.

2. *The Olympics have the highest average cost overrun of any type of megaproject*, at 172 percent in real terms. To compare, Flyvbjerg et al. (2002) found average cost overruns in major transportation investments of 20 percent for roads, 34 percent for large bridges and tunnels, and 45 percent for rail; Ansar et al. (2014) found 90 percent overrun for megadams; and Budzier and Flyvbjerg (2011) 107 percent for major IT investments, all in real terms and measured in the same manner (see Table 4). The high cost overrun for the Games may be related to the fixed deadline for delivery: the opening date cannot be moved. Therefore, when problems arise there can be no trade-off between schedule and cost, as is common for other megaprojects. All that managers can do at the Olympics is to allocate more money, which is what happens. This is the Blank Check Syndrome, again.

3. *The high average cost overrun for the Games, combined with the existence of extreme values, should be cause for caution for anyone considering to host the Games*, and especially small or fragile economies with little capacity to absorb escalating costs and related debt. Even a small risk of a 50+ percent cost overrun on a multi-billion dollar investment should concern government officials and taxpayers when a guarantee to cover cost overrun is imposed, because such overrun may have fiscal implications for decades to come, as happened with Montreal where it took 30 years to pay off the debt incurred by the 720 percent cost overrun on the 1976 Summer Games (Vigor et al. 2004: 18), and Athens 2004 where Olympic cost overruns and related debt exacerbated the 2007-17 financial and economic crises, as mentioned above (Flyvbjerg 2011).

[Table 4 app. here]

**The Power Law of Olympic Cost**

To identify more specifically the type of risk involved in bidding for the Olympics, we fitted different theoretical statistical distributions to the data. We did this for data on cost overrun, bid cost, outturn cost, and cost per athlete, all presented above. Here we focus on cost overrun, because this is the most critical variable for potential host cities and nations in deciding whether they can afford to host the Games and would be likely to stay on budget or not. It should be mentioned, however, that results for the other variables are similar.

We tried all possible theoretical frequency distributions and found that only fat-tailed distributions fit the data, i.e., distributions with many extreme values. Thin-tailed distributions can be rejected, whereas fat-tailed ones cannot. Power law and lognormal distributions were found to best fit the data (see Figure 4).[6]

[Fig. 4 app. here]

Power laws show potentially infinite variance and a volatile or nonexistent mean. As a consequence of the fat-tailedness of the data we cannot reliably establish the presence of time trends or compare means between subgroups of data, e.g., Summer Games v. Winter Games; the extreme values in the fat tail affect the results too much for this type of approach to be feasible. In essence, with fat tails the law of large numbers is not fast enough to give a reliable idea of the mean. This problem is exacerbated by the fact that each of the Summer and Winter Games are held only every four years, so by the nature of their setup a large number of Games are not and cannot be available for study.

We examined in detail three models that fit the data on cost overrun particularly well, one lognormal and two power laws. First, the maximum likelihood estimator (MLE) for the lognormal model gives $\mu = 0.85$ and $\sigma = 0.533$. While conventional thinking is that a lognormal distribution is thin-tailed, because it has all the moments (mean, variance, etc.), it actually behaves like a power law at a $\sigma > 0.4$ (Taleb 2020: 139 ff.), which is the case for our Olympic costs dataset. The Kurtosis for the lognormal model is 7.10. The probability of costs being three times estimates is 22.4 percent.

Second, the maximum likelihood estimator for the first power law – the Pareto 1, a simplified power law – finds $\lambda = 1.015$ (the minimum value in the sample) and the power law coefficient $\alpha = 1.199$. The advantage of the simplified power law is the estimation of a single parameter, $\alpha$.

Finally, the second power law uses three parameters and for this model we find $\alpha \approx 1.6$, which indicates thinner, but still fat, tails. We observe that $1 < \alpha < 2$ for both models, which means that the first moment (mean) is finite and can therefore be estimated, whereas the second moment (variance) is infinite and therefore cannot. If $\alpha$ were smaller than one the tail would be so fat that both mean and variance would be infinite and neither could be estimated.

The controversy over the power-law fittings done by Barabási et al. (2005) highlights that careful consideration must be observed when fitting and comparing fat-tailed distributions so as to not overstate the case (Stouffer et al. 2005, Barabási et al. 2005, Shalizi 2005, Clauset 2005). To guard against the pitfall of overstatement, we tested how the power-law models compare with the lognormal model in fitting the data. For this, we used Vuong's test, which is a likelihood ratio test for model selection using the Kullback-Leibler criteria. The null hypothesis is that both classes of distributions (power law and lognormal) are equally far from (or near to) the true distribution. The test statistic, R, is the ratio of the log-likelihoods of the data between the two competing models. The sign of R indicates which model is better. The size of R indicates the likelihood of the null hypothesis being true or false, with R going to ±infinity with probability 1 if the null hypothesis is false, indicating that one type of distribution is closer to the true distribution than the other. Using this test, we found R = -1.193668 with p = 0.0851, which favors the power law fit, however not at a level that allows us to reject the null.

Following Clauset et al. (2009) we then used their iterative approach to search for values of λ that would improve the power law fit. We found λ = 1.494919 and α = 1.711926. The MLE fit of the lognormal distribution for the same λ gives lnN(0.5438; 0.7124). Vuong's likelihood ratio test now results in R = -0.251 with p = 0.802. At this higher starting point of the tail the model comparison can again not reject the null that either distribution (power law and lognormal) fits the data.

We conclude that both the power law distribution and the lognormal distribution fit the data on Olympic cost overrun and that both distributions are fat tailed. The findings translate into the following fast-and-frugal, practical heuristic for cities and nations trying to decide whether to host the Games or not:

> Heuristic no. 1: *Can we afford and accept a 20 percent risk of a three-fold increase or higher in cost in real terms on the multi-billion-dollar expenditure for the Olympics? If the answer to this question is yes, then proceed and become a host; if the answer is no, walk away.*

**The Law of Regression to the Tail**

Figure 5 shows eight states of randomness, from zero ("degenerate") to infinite ("α ≤ 1"). Thin-tailed distributions, with low randomness, are at the bottom of the figure. Fat-tailed distributions, with high randomness, at the top. The eight states are somewhat similar to Mandelbrot's (1997) classic categories of mild, slow, and wild randomness, but are statistically more detailed with regards to the implications for modelling risks.

[Figure 5 app. here]

We saw above that Olympic cost overruns follow fat-tailed distributions, specifically lognormal and power-law distributions with an alpha-value of 1.2 to 1.7, i.e., smaller than 2 and larger than 1. This is a highly significant finding to anyone who wants to understand how Olympic costs work, and especially to anyone considering to host the Games concerned about affordability and financial risks. Comparing the established alpha-values with Figure 5, we see that Olympic cost overruns belong to the second-most extreme category in terms of randomness, here called "Lévy-Stable" after Lévy processes in probability theory (1 < α < 2). Distributions in this category are characterized by infinite variance and finite mean, whereas for the most extreme category (α ≤ 1) variance and mean are both infinite, i.e., non-existent.

With this finding we arrive at a basic empirical explanation of Olympic cost blowouts that has not been uncovered before. Such events are not just the unfortunate, happenstance incidents they appear to be, that are regrettable but will hopefully be avoided in the future, with more awareness and better luck. Instead, Olympic cost blowouts are systematic, ruled by a power law that will strike again and again, with more and more disastrous results. Following power-law logic, it is just a matter of time until an event, here cost overrun, will occur that is even more extreme, with a larger cost overrun, than the most extreme event to date. We call this phenomenon "regression to the tail" and contrast it with "regression to the mean."

Sir Francis Galton coined the term regression to the mean – or "regression towards mediocrity," as he originally called it. It is now a widely used concept in statistics, describing how measurements of a sample mean will tend towards the population mean when done in sufficient numbers, although there may be large variations in individual measurements. Galton illustrated his principle by the example that parents who are tall tend to have children who grow up to be shorter than their parents, closer to the mean of the population, and vice versa for short parents. In another example, pilots who performed well on recent flights tended to perform less well on later flights, closer to the mean of performance over many flights. This was not because the pilots' skills had deteriorated, but because their recent good performance was due not to an improvement of skills but to lucky combinations of random events. Regression to the mean has been proven mathematically for many types of statistics and is highly useful for risk assessment and management, in, e.g., health, safety, insurance, casinos, and factories.

But regression to the mean presupposes that a population mean exists. For some random events of great social consequence this is not the case. Size-distributions of earthquakes, wars, floods, cybercrime, bankruptcies, and IT procurement, e.g., have no population mean, or the mean is ill defined due to infinite variance. In other words, mean and/or variance do not exist. Regression to the mean is a meaningless concept for such distributions, whereas regression to the tail is meaningful and consequential. Regression to the tail applies to any distribution with non-vanishing probability density towards infinity. The frequency of new extremes and how much they exceed previous records is decisive for how fat tailed a distribution will be, that is, whether it will have infinite variance and mean. Above a certain frequency and size of extremes, the mean increases with more events measured, with the mean eventually approaching infinity. In this case, regression to the mean means regression to infinity, i.e., a non-existent mean. Deep disasters – e.g., earthquakes, tsunamis, pandemics, and wars – tend to follow this type of distribution. We call this phenomenon – extreme events recurring in the tail, with events more extreme than the most extreme so far – "the law of regression to the tail" (Flyvbjerg forthcoming). We show that cost overrun for the Olympic Games follow this law.

Table 5 compares the alpha-values for the Olympic Games with alpha-values for other events that follow power-law distributions. We see that with alpha-values between 1 and 2, indicating infinite variance, Olympic cost overruns fall in the same category as terrorist attacks, forest fires, floods, and bankruptcies in terms of fat-tailedness.

[Table 5 app. here]

A further consequence of power-law behavior and regression to the tail, is that the sample mean cannot be trusted: even a single extreme value in the fat tails may significantly alter the mean. However, knowing that the power law distribution and the lognormal distribution fit the Olympic cost data, we may use this knowledge to estimate the true (population) mean of the distributions, following Taleb and Cirillo (2015).[7] Table 6 shows the results. The sample mean, as we saw above, is 2.72, equivalent to an average cost overrun of 172 percent. The true mean, based on the lognormal distribution, is 2.69, i.e., practically the same as the sample mean. Based on the two Pareto distributions, we find a true mean of 8.46 and 3.59, respectively, i.e., substantially higher than the sample mean. The best estimate of the true mean of cost overrun in the Olympics is therefore 172-746 percent.

This finding gives us further detail to heuristic no. 1 above and translates into a second fast-and-frugal heuristic that cities and nations may find useful when faced with deciding whether to host the Games:

*Heuristic no. 2: Can we afford and accept an expected cost overrun in the range of 170-750 percent in real terms on the multi-billion-dollar expenditure for the Olympics, with substantial risk of further overrun above this range? If the answer to this question is yes, then proceed and become a host; if the answer is no, walk away, or find effective ways to "cut the tail," i.e., reduce tail risk.*

[Table 6 app. here]

**Explaining the Olympic Power Law**

What is the mechanism that generates the power-law distribution for Olympic cost and cost overrun? What are the implications? And would it be possible to mitigate the impacts of the power law? We address these questions below.

At the most basic level, power laws are generated by non-linear amplification of random events, including infinite amplification (Farmer and Geanakoplos 2008).[8] In other words, when the value of an underlying variable changes, say the scope of the Games, then the resulting output – say, cost – does not change linearly, but instead depends on the derivative. The change is accelerated, so to speak. Geometrically, the relationship between variables is curved instead of linear. The degree of curvature is called convexity. The more curved, the more convexity, and the more amplification of random events. At the level of root causes, convexity is the primary mechanism that generates power laws, including the Olympic cost power law. Convexities are known to lead to serious financial fragilities (Taleb 2012). Convexity appears to be severe for Olympic cost overruns, as evidenced by the alpha-values shown in Table 5, documenting infinite variance. We find six reasons for this.

First, hosting the Games is a particularly difficult decision to reverse. This means that when scope and costs begin to escalate – as they have for every Games on record – hosts generally do not have the option of walking away, as they do with most other investments, even should they think this would be the best course of action. In fact, Denver is the only host city to ever have abandoned the Games, in 1972, after winning the bid to host them.[9] Instead, lock-in is near-absolute and hosts are forced to throw good money after bad, which is the type of behavior that leads to strong convexity in spending.

Second, not only is it difficult to reverse the decision to host the Games, there is also no option to save on costs by trading off budget against schedule, because the timetable for the Olympics is set in stone. The bid is won seven years in advance of the opening ceremony, the date of which was decided even earlier and cannot be moved. For other types of megaprojects, trading off budget against schedule is a common and often effective mechanism to dampen cost escalation. For the Games, this mechanism is not available, again reinforcing convexity and power-law outcomes.

Third, there is the legally binding obligation, mentioned above, that the host must cover possible cost overruns. This means that the IOC has no incentive to curb cost overruns, but quite the opposite, as the IOC focuses on revenues, from which their profits derive, and some costs drive revenues and thus IOC profits.[10] The host, on the other hand, has no choice but to spend more, whenever needed, whether they like it or not. This is the Blank Check Syndrome mentioned above. The blank check is, in and of itself, an amplification mechanism that generates convexity and power law outcomes for Olympic cost and cost overrun.[11] Things do not need to be like this. The IOC could choose to lead on costs instead of self-servingly focusing on revenues. But so far they do not.

Together, the first three points make for a clear case of strong convexity and go a long way in explaining why Olympic costs are fat tailed. The way the Games are set up, including their contracts and incentives, virtually guarantees that strong convexity and power-law outcomes ensue. But there is more.

Fourth, tight constraints on scope and quality in the delivery of investments are known to be an additional driver of convexity and power law outcomes (Carlson and Doyle 1999). Together with the immovable opening date for the Games, mentioned above, a program scope that is rigorously defined by the many sports and events hosted, and design standards that are set in detail by the IOC and individual sports associations are examples of tight and non-negotiable constraints that set the Olympics apart from more conventional megaprojects where trade-offs between budget, schedule, scope, and quality have wider margins. For the Olympics such margins are zero or close to zero. For example, the delivery authority does not get to negotiate the standards for running the 100 meter or for bob sleighing. They are givens, just like the opening date, as far as delivery is concerned. This means that staging the Games can be conceived as a highly optimized system with budget uncertainty at the macro level, while constraints are exceptionally tight at the micro level. Such systems have been demonstrated to be fragile to design flaws and perturbations and subject to convexity and power law outcomes. Moreover, constraints and their effects are exacerbated by size, which the IOC and hosts should keep in mind as the Olympics grow ever bigger (Taleb 2012: 279-80).

In conventional investment management, constraints are softened by the use of contingencies. So too with the Olympics where bid budgets typically include reserves. For example, the bid for London 2012 included a 4.3 percent contingency. This proved sorely inadequate in the face of a 76 percent cost overrun in real terms, i.e., an amplification 18 times higher than the contingency allowed for. Ten years later, with the bid for Beijing 2022, not much seemed to have changed. Here the budget included a 9.1 percent contingency for unanticipated expenses, which, according to the IOC "is in line with the level of risk and the contingency for previous Games" (IOC 2015: 75). Whether in line with previous practice or not, such contingencies are sorely inadequate, and Table 3 demonstrates they are not in line with the level of risk for previous Games, as falsely claimed by the IOC. More than for any other type of megaproject, the contingency mechanism – which is crucial to effective delivery – fails for the Olympics. When contingencies run out, as they have done for every Olympics on record, such failure typically leads to what we call the "vicious circle of cost overrun," where top management gets distracted from delivery, because they are now forced to focus on pressing issues of negative media coverage about lack of funds, reputational damage, and fund raising aimed at closing the contingency gap. As a consequence, delivery suffers, which leads to further cost overruns, which lead to further distractions, etc. The vicious circle is caused in part by the exceptionally tight constraints that apply to delivery of the Games and the high levels of contingency that are needed, but not available, to soften such constraints. Again the consequence is higher convexity and power law outcomes.[12]

Fifth, the longer the planning horizon the higher the variance of random variables and the more opportunities for random events to happen that may get amplified to power law outcomes (Makridakis and Taleb 2009, Farmer and Geanakopolos 2008). Time is like a window. The longer the duration, the larger the window, and the greater the risk of a big, fat black swan flying through it, which tends to happen for the Olympics, with its overincidence of extreme values documented above. Some of these black swans may be deadly, in the sense that they kill plans altogether, as happened for the Tokyo 2020 Olympics when the covid-19 pandemic made it impossible to host the Games in 2020 as planned. This resulted in billions of dollars of further cost overruns on top of billions already incurred, making the fat tails for Olympic cost even fatter.

By design, staging the Olympics comes with a long planning horizon, specifically seven to eleven years.[13] This is the length of an average business cycle in most nations. It should therefore come as no surprise that the price of labor and materials, inflation, interest rates, exchange rates, etc. may change significantly over a period this long and impact cost and cost overrun. Cities and nations typically bid for the Games when the economy is thriving, with the consequence that more often than not the business cycle has reversed to lower growth when the opening date arrives seven to eleven years later, as we saw in Figure 1.

Furthermore, scope changes generated by random events will be the more likely to happen the longer the planning horizon, e.g., terrorist attacks that push up security standards and costs at the Games, as has happened repeatedly. Interestingly, the severity of terrorist attacks, measured by number of deaths, follow a power law distribution, just like Olympic costs. This is an example of one power law (number of deaths from terrorist attacks) directly driving another power law (cost and cost overrun at the Games). This is archetypical amplification and strong convexity at work.

In addition to the risk of underestimating the true variability of forecasts, long planning horizons fundamentally change the nature of risk encountered. Power laws arise when clustered volatility and correlated random variables exists in a time series, as has been observed for financial data (Mandelbrot 1963, Gabaix 2009). Such time series are dominated by random jumps instead of the smooth random walks often assumed by analysts and forecasters (Mandelbrot 1963). And random jumps – e.g., a sudden increase in the price of steel, a main ingredient of the Olympics, or a jump in security costs triggered by a terrorist attack – lead to power law outcomes. The longer the planning horizon the more likely that random jumps will happen.

Existing forecasting techniques cannot deal with outcomes like these, if at all, beyond planning horizons of approximately one year (Blair et al. 1993, Gabaix 2009, Beran 2013, Tetlock and Gardner 2015). In hydrology, Mandelbrot and Wallis (1968) called the rare instances of extreme precipitation that occur over the long term – often after long periods of low precipitation – the "Noah effect." The coining of the term was inspired by the empirical observations of the fluctuations of the water level in the Nile River by Harold Edwin Hurst, but the effect works beyond hydrology. Based on his observations, Hurst increased the height of the planned Aswan High Dam far beyond conventional forecasts, and the dam was designed and built following his recommendations. Hurst did not try to forecast the exact maximum level of water in the dam basin, which would have been in vain, and would likely have led to underestimation. Instead he increased the height of the dam wall to take into account extreme values far beyond his observations. This is exactly what power laws dictate as the right course of action, because according to the power law it is only a matter of time until an event occurs that is more extreme than the most extreme event to date. Unlike Hurst, cost forecasters at the Olympics fail to understand the power-law nature of the phenomenon at hand and therefore fail to increase their safety margins, with the consequence that they have underestimated costs for every Olympics on record.

Sixth and finally, the problems described above are compounded by a phenomenon we call the Eternal Beginner Syndrome. If, perversely, one wanted to make it as difficult as possible to deliver a megaproject to budget and schedule, then one would (a) make sure that those responsible for delivery had never delivered this type of project before, and (b) place the project in a location that had never seen such a project, or at least not for the past few decades so that any lessons previously learned would be either obsolete or forgotten. This, unfortunately, is a fairly accurate description of the situation for the Olympics, as they move from nation to nation and city to city, forcing hosts into a role of "eternal beginners."[14] It is also a further explanation of why the Games hold the record for the highest cost overrun of any type of megaproject.

Inexperienced beginners are more prone than experienced experts to underestimate challenges and are less well suited in dealing with unexpected events when they happen. This means that such events spin out of control and amplify more easily for beginners than for experienced experts, which again contribute to convexity and the power-law outcomes we see in the data. A mistake made by eternal beginners is to assume that delivering the Olympics is like delivering a scaled-up but otherwise conventional construction program. This emphatically is not the case. Rio 2016, for example, hosted 42 sports with 306 events in 32 venues. Tight coupling of the deliverables, fixed deadlines, and inadequate budget contingencies form a system that amplifies the impacts of random adverse events (Dooley 1999). The eternal beginner lacks experience in what this kind of scale and constraints mean in terms of increased delivery risks and therefore underestimates these. The lack of experience is aggravated by the fact that conventional approaches to estimating and containing risk do not work in this kind of system, as argued above.

In sum, we find: (a) convexity is the root cause of the power-law nature and extreme randomness of cost and cost overrun for the Olympics; (b) convexity is strong for the Games, documented by alpha-values smaller than 2, indicating infinite variance; and (c) convexity at the Games is driven by irreversibility, fixed time schedules, misaligned incentives, tight coupling, long planning horizons, and the Eternal Beginner Syndrome.

**Six Steps to Better Games Management**

Power-law fat tails – or extreme randomness – is the most challenging type of risk to manage, because it is maximal, unpredictable, and difficult to protect against via conventional methods. However, this is the type of risk that the Olympics face in terms of cost, as shown above. What can be done to manage this risk intelligently?

First, and most importantly, *the IOC and potential hosts must understand the existence of fat tails* as a matter of fact, i.e., that hosting the Games is extremely risky business in terms of cost. If they do not understand the risk, and its particular power-law nature, they cannot hope to effectively protect themselves against it. Today, there is some understanding of risk with the IOC and hosts, but nothing that reflects the real risks. Instead of extreme randomness, the IOC assumes low randomness when it states that a 9.1 percent contingency for unanticipated expenses is in line with the level of risk for previous Games, as we saw above. This number is glaringly insufficient when compared with actual cost overrun in the most recent Games, which was 352 percent for Rio 2016, 289 percent for Sochi 2014, and 76 percent for London 2012, or when compared with the average cost overrun for all Games since 1960, which is 172 percent. Either the IOC is deluded about the real cost risk when it insists that a 9.1 percent contingency is sufficient, or the Committee deliberately overlooks the uncomfortable facts. In either case, host cities and nations are misled, and as eternal beginners it is difficult for them to protect themselves against such misinformation. They understandably do not know what the real numbers are, because they have no experience in delivering the Games. Independent review of any cost and contingency forecast is therefore a must, including for estimates from the IOC. As said, hosts must understand the risk to be able to protect themselves against it. Such understanding is therefore a necessary first step for mitigating cost risk at the Games, and the IOC should be held accountable for misinforming hosts about the real risks under rules similar to those in corporate governance that make it unlawful for corporations to deliberately or recklessly misinform shareholders and investors.

Second, once the real risks are understood it becomes immediately clear that *larger cost contingencies are needed for the Games*. Reference class forecasting, based on Kahneman and Tversky's (1979) work on prediction under uncertainty, has been shown to produce the most reliable estimates of any forecasting method, mainly because it eliminates human bias and takes into account unknown unknowns (Kahneman 2011: 251, Flyvbjerg 2008, Batselier and Vanhoucke 2016 and 2017, Chang et al. 2016, Horne 2007). Further de-biasing should be carried out, following Kahneman et al. (2011) and Flyvbjerg (2013). Based on the dataset presented above, reference class forecasting would produce a significantly more realistic estimate of the necessary cost contingencies for the Olympics than the numbers put forward by the IOC. More realistic contingencies would have the additional advantage of softening the tight constraints identified above as a driver of power-law outcomes at the Games, which in turn would help drive down cost blowouts and costs. In this manner, more realistic contingencies could help make a first clip in the fat tail of Olympic costs. However, larger contingencies alone will not solve the problem. Cost risks must also be actively managed down.

Third, *the IOC should have skin in the game* as regards cost, i.e., it should hold some of the cost risk that arise from staging the Games (Taleb and Sandis 2014). The IOC sets the agenda, defines the specs, and has ultimate decision making power over the Games. Nevertheless, the IOC holds none of the cost risk involved. As a result there is little alignment between incentives to keep costs down and making decisions about cost, which is one reason costs explode at the Games, and will keep exploding. For any other type of megaproject such massively misaligned incentives would be unheard of. In order to change this state of affairs, we suggest the IOC is made to cover from its own funds minimum 10 percent of any cost overrun for the Games, to be paid on an annual basis as overruns happen. This would give the IOC the motivation it lacks today to effectively manage costs down and thus help reduce the Blank Check Syndrome. Cities and nations should refuse to host the Games unless the IOC agrees to do this. Lack of such agreement would be a clear sign that the IOC does not take cost control seriously. We further suggest that antitrust bodies take a look at the IOC, which today is an unregulated monopoly, and consider regulating it for better performance in accordance with antitrust law.

Fourth, anything that can be done to *shorten the seven-year delivery phase for the Games* should be considered. The longer the delivery, the higher the risk, other things being equal. For many Games, not much happens the first 2-3 years after winning the bid, which indicates that faster delivery would be possible, as it was before 1976. Faster delivery may be supported by a more standardized and modularized approach to delivery, without the need to reinvent the wheel at every Games, and by using existing facilities as much as possible. Here it is encouraging to see that the IOC has decided to consider "turnkey solutions" for OCOGs in areas that require highly specific Olympic expertise (IOC 2014: 15). Standardized turnkey solutions should be developed as far as possible to help hosts reduce costs. Finally, a much more ambitious goal could be set for schedule and cost, to drive innovation at the Games, for instance: "*Games delivered at half the cost, twice as fast, with zero cost overrun.*" We suggest that going forward the IOC adopts this slogan as one of its goals. That would show true ambition and willingness to innovate regarding cost control. Today's budgets and schedules are so bloated that this goal would not be unrealistic for a professional and experienced delivery organization. Unfortunately, as a monopoly that answers to no one the IOC is unlikely to innovate unless it is forced to do so from the outside.

Fifth, to directly *tackle the Eternal Beginner Syndrome* proposals have been put forward to host the Games in one or a few permanent locations – e.g., Athens – or, alternatively, that two successive Games should be given to the same host, so facilities could be used twice (Short 2015, Baade and Matheson 2016). As a further variation on this theme, Games could be spread geographically with different events going to different cities, but with each event having a more or less permanent home, say track and field in Los Angeles, tennis in London, equestrian events in Hong Kong, etc. This could be combined with a permanent and professional delivery authority, responsible for staging the Games every time and accumulating experience in one place in order to secure effective learning and build what has been called a "high-reliability organization" for delivering the Games, something that has been sorely missing so far in the history of the Games (Roberts 1990).

Finally, and perhaps most effectively, *prospective host cities could mitigate their risk by simply walking away from the Games*. Indeed, this has become a preferred strategy for many cities. Over the past 20 years the number of applicant and candidate cities have fallen drastically, from a dozen to a few (Zimbalist 2015: 24; Lauermann and Vogelpohl 2017). Most recently, Barcelona, Boston, Budapest, Davos, Hamburg, Krakow, Munich, Oslo, Rome, Stockholm, and Toronto have made highly visible exits from the bid process. For the 2022 Winter Olympics, the IOC ended up with just two potential hosts, namely Almaty in Kazakhstan and Beijing, with the latter – not known as one of the world's great winter sports resorts – winning the bid. For the 2024 Summer Olympics the same happened, with Paris and Los Angeles as the only potential hosts, after three other cities pulled out. To protect itself from further humiliation with the 2028 Summer Games, the IOC broke precedent and selected the next two Summer Games hosts simultaneously by negotiating a deal with Paris and Los Angeles, according to which Paris would win the bid for the 2024 Games and the 2028 Games would go to Los Angeles without a bid round. Finally, for the 2026 Winter Olympics the IOC had to extend the invitation phase after no city had bid for the Games well into 2017.

The exodus of candidate cities has been acutely embarrassing to the IOC and has caused reputational damage to the Olympic brand. Cities have explicitly and vocally cited extravagant costs and cost overruns as a main reason for exiting the bid process. With Agenda 2020, which is being touted by the IOC as protecting the interests of host cities, the IOC has made a first attempt to address the exodus, and to protect its brand and product (IOC 2014, 2017). But the initiative, though welcome, does not address the root causes and looks like too little too late. Time will tell, with Tokyo 2020 as the first test case, if it is held as planned in 2021. Our guess is that more reform will prove necessary to stop the exodus. As rightly observed by Los Angeles mayor Eric Garcetti, "Most cities, unless you have a government that's willing to go into debt or pay the subsidy of what this costs, most cities will never say yes to the Olympics again unless they find the right model," with the "right model" being defined by significantly lower costs (Ford and Drehs 2017). Garcetti said this before the covid-19 pandemic and its large extra debt burdens on governments. Post-covid-19, the appetite and ability for governments to go into further debt or pay a subsidy to finance the Olympic Games will likely be low and the pressure to keep costs down will likely be high. Perhaps this can finally get the IOC to take Olympic costs and cost overruns seriously and try to manage them down in collaboration with host cities. The novel explanations of Olympic costs and cost risks uncovered in the present paper shows that the IOC and host cities would have every reason to do so and provides the evidence and guidelines for how to accomplish the task.

**Acknowledgments**

The authors wish to thank Nassim Nicholas Taleb for helping develop key ideas of the paper and for carrying out the statistical analysis involved in fitting the theoretical statistical distributions to the data. Allison Stewart helped collect the data for the Games 1960-2012 and co-authored Flyvbjerg and Stewart (2012), on which parts of the current paper builds. Finally, the authors wish to thank the review team at the journal for highly useful comments and suggestions of improvements to an earlier version of the paper.

**Annex 1: Previous Studies**

Interest in cost and cost overrun of the Games has been high since the establishment of the modern Olympics in 1896. As long ago as 1911 Baron Pierre de Coubertin, founder of the IOC and the modern Games, referred to "the often exaggerated expenses incurred for the most recent Olympiads" (Coubertin, 1911), and in 1973 Jean Drapeau, the mayor of Montreal, infamously stated, "The Montreal Olympics can no more have a deficit, than a man can have a baby," which caused some peculiar cartoons in Canadian media when the Montreal Games incurred a large deficit due to the biggest-ever Olympic cost overrun, at 720 percent (CBC 2006). Drapeau was wrong, and problems with cost and cost overrun are as prevalent today as they were in his time, and in Coubertin's before him.

Despite substantial interest in the cost of the Games, however, attempts to systematically evaluate such cost are few, perhaps because valid and reliable data that allow comparative analysis are difficult to come by (Chappelet 2002, Kasimati 2003, Essex and Chalkley 2004, Preuss 2004, Zimbalist 2015, Baade and Matheson 2016). Instead, the attempts that exist have typically focused on a specific or a few Games (Brunet 1995, Bondonio and Campaniello 2006, Müller and Gaffney 2018). Another strain of research has focused on whether the Games present a financially viable investment from the perspective of cost-benefit analysis (Zimbalist 2015: 31-70). But what to measure when determining the costs and benefits of the Games is open to debate and has varied widely between studies, again making it difficult to compare results across games and studies. In particular, legacy benefits described in the bid are often intangible, and as such pose a difficulty in ex-post evaluations. The benefits of increased tourism revenue, jobs created, or national pride are hugely varied and similarly difficult to quantify and compare. Costs can also be hard to determine; for example, one could argue that if hotels in the host city have invested in renovations, and benefits of increased tourist revenues to those hotels are included in the analysis, then these costs should also be included in any accounting, but they rarely are. Finally, the percentage of work that an employee in an outlying city spends on Games-related work would be exceptionally difficult to estimate.

Preuss (2004) contains the perhaps most comprehensive multi-Games economic analysis to date, looking at the final costs and revenues of the Summer Olympics from 1972 to 2008. Preuss finds that since 1972 every Organizing Committee of the Olympic Games (OCOG), which leads the planning of the Games in the host city, has produced a positive benefit as compared to cost, but only when investments are removed from OCOG budgets. This restricts the analysis narrowly to only OCOG activities, which typically represent a fairly small portion of the overall Olympic cost and therefore, we argue, also denotes too limited a view for true cost-benefit analysis. Further, other authors disagree with Preuss' findings, and have suggested that the net economic benefits of the Games are negligible at best, and are rarely offset by either revenue or increases in tourism and business (Malfas, Theodoraki, and Houlihan 2004; Billings and Holladay 2012; von Rekowsky 2013; Goldblatt 2016: 392-93). Furthermore, none of these studies have systematically compared projected cost to final cost, which is a problem, because evidence from other types of megaprojects show that cost overruns may, and often do, singlehandedly cause positive projected net benefits to become negative (Flyvbjerg 2016; Ansar et al. 2017). Taking the total body of knowledge into account, a recent study of the economics of the Olympics, published in *Journal of Economic Perspectives*, found that "the overwhelming conclusion is that in most cases the Olympics are a money-losing proposition for host cities" (Baade and Matheson 2016: 202).

But there may be other legitimate reasons for hosting the Olympics than mere money-making, for instance national pride or throwing the biggest party on the planet. Whatever the reason for hosting the Games, hosts should know the financial and economic risks involved, before making the decision.

In sum, we find for previous academic research on cost and cost overrun for the Olympic Games:

1. Earlier attempts to systematically evaluate cost and cost overrun in the Games are few;
2. Such attempts that exist are often focused on a specific Games or are small-sample research;
3. Earlier research on the cost of the Games has focused on cost-benefit analysis, with debatable delimitations of costs and benefits making it difficult to systematically compare results across studies and Games;
4. Existing evidence indicate that benefits generally do not outweigh costs for host cities and nations.

Flyvbjerg and Stewart (2012) is the first study to consistently compare cost and cost overrun for a large number of Olympic Games. That study took its inspiration in comparative research more broadly, looking at megaprojects, and used a method for measuring cost and cost overrun that is the international standard in this research field. The goal was to bring the same rigor and consistency to the study of Olympic costs as that found in megaproject cost scholarship.

Previous research has established clearly that hosting the Olympics is costly and financially risky. There is a gap in previous research, however, regarding the specific nature of the cost risks faced by cities and nations deciding whether to host the Games or not. The present paper aims to fill this gap, by fitting different theoretical statistical distributions to Olympic cost data, as described in the main text.

**Annex 2: Data and Methodology**

Olympic bidding for a specific Games typically begins with a country's National Olympic Committee making a call for expressions of interest from prospective host cities eleven years before the Games in question. Interested cities then compete to become their country's favorite, which is decided nine years prior to the Games. Finally, among the favorite cities, which are known as "applicants," the IOC typically chooses three to five "candidate" cities, which enter the concluding competition to host the Games, decided seven years before the Games take place, with one city declared winner of the bid.

In the competition to win, cities pitch their ideas to the IOC for how best to host the world's biggest sporting event and how to generate significant urban development in the process (Andranovich, Burbank, and Heying 2001). To demonstrate their ability to achieve these goals, bidding cities are required by the IOC to develop detailed plans in the form of so-called Candidature Files that are submitted to the IOC as part of the competition to host. The Candidature Files, or "bid books" as they are more commonly known, form part of the basis of the IOC's decision for the next host city.

One of the requirements for the bid book is that it includes a budget that details the expected investment by the host, in addition to a budget for expected revenues (IOC 2004). In their bid book, governments of candidate cities and nations are also required by the IOC to provide guarantees to "ensure the financing of all major capital infrastructure investments required to deliver the Olympic Games" and "cover a potential economic shortfall of the OCOG [Organizing Committee of the Olympic Games]" (*ibid*: 93).

The Candidature File is a legally binding agreement, which states to citizens, decision makers, and the IOC how much it will cost to host the Games. As such the Candidature File represents the baseline from which future cost and cost overrun should be measured. If cost overrun later turns out to be zero, then decision makers made a well-informed decision in the sense that what they were told the Games would cost is what they actually ended up costing, so they had the correct information to make their decision. If cost overrun is significantly higher than zero, then the decision was misinformed in the sense that it was based on an unrealistically low estimate of cost. However, such measurement of cost against a consistent and relevant baseline is rarely done for Olympic costs. New budgets are typically developed after the Games were awarded, which are often very different to those presented at the bidding stage (Jennings 2012). These new budgets are then used as new baselines, rendering measurement of cost overrun inconsistent and misleading both within and between Games. Using later baselines typically makes cost overruns look smaller and this is a strong incentive for using them, as in the case for London 2012 mentioned in the introduction to the main text. New budgets continue to evolve over the course of the seven years of planning for the Games, until the final actual cost is perhaps presented, often several years after the Games' completion – if at all, as we will see.

Our objective was to measure cost and cost overrun for the Games in a consistent and relevant manner. We therefore searched for valid and reliable bid book and outturn cost data for both Summer and Winter Games, starting with the Rome 1960 Summer Games and the Squaw Valley 1960 Winter Games, and continuing until the most recent Winter and Summer Games.

Costs for hosting the Games fall into the following three categories, established by the IOC:

1. *Operational costs* incurred by the Organizing Committee for the purpose of "staging" the Games. These include workforce, technology, transportation, administration, security, catering, ceremonies, and medical services. They are the variable costs of staging the Games and are formally called "OCOG costs" by the IOC.
2. *Direct capital costs* incurred by the host city or country or private investors to build the competition venues, Olympic village(s), international broadcast center, and media and press center, which are required to host the Games. These are the direct capital costs of hosting the Games and are formally called "non-OCOG direct costs."
3. *Indirect capital costs*, for instance for road, rail, or airport infrastructure, or for hotel upgrades or other business investment incurred in preparation for the Games but not directly related to staging the Games. These are wider capital costs and are formally called "non-OCOG indirect costs."

The first two items constitute the *sports-related costs* of the Games and are covered in the present analysis. Non-OCOG indirect costs have been omitted, because (1) data on such costs are rare, (2) where data are available, their validity and reliability typically do not live up to the standards of academic research, and (3) even where valid and reliable data exist, they are often less comparable across cities and nations than sports-related costs, because there is a much larger element of arbitrariness in what is included in indirect costs than in what is included in sports-related costs; plus many indirect costs cover expenditures that would have been incurred even without the Games, although perhaps at a later time. It should be remembered, however, that the indirect costs are often higher than the direct costs. Baade and Matheson (2016: 205) found that for seven Games for which they could obtain data for both sports infrastructure and general infrastructure, in all cases the cost of general infrastructure was higher than the cost of sports infrastructure, sometimes several times higher. For example, for Barcelona 1992, the cost of general infrastructure was eight times that of sports infrastructure; for Vancouver 2010 five times higher. As developing nations increasingly bid for the Games, costs for general infrastructure are bound to get even higher, because emerging economies typically have inadequate transportation, communications, energy, water, hospitality, and other infrastructure that must be upgraded before the Games can be hosted (Zimbalist 2015: 2, Fletcher and Dowse 2017).

For measuring final outturn sports-related cost, data were available for 25 out of the 30 Games 1960-2016, or for 83 percent of Games. For measuring cost overrun, which involves comparing estimated bid cost with final outturn cost, data were available for 19 of 30 Games. For the remaining 11 Games, valid and reliable data have not been reported that would make it possible to establish cost overrun for these Games. This is an interesting research result in its own right, because it means – incredible as it may sound – that for more than a third of the Games since 1960 no one seems to know what estimated and actual costs were. In addition to being a powerful indictment of Olympic bidding and of the opacity of local Olympic boosterism and IOC decision making, such ignorance hampers learning regarding how to develop more reliable budgets for the Games. From a rational point of view, learning would appear to be a self-evident objective for billion-dollar events like the Games, but often that is not the case. For some Games, hiding costs and cost overruns seems to have been more important, for whatever reason. Nevertheless, 19 out of 30 Games is 63 percent of all possible Games for the 60 years under consideration, which we deem sufficient for producing interesting and valid results.

We measured costs in both nominal and real (adjusted for inflation) terms, and in both local currencies and US dollars. We followed international convention and made all comparisons across time and geographies in real terms, to ensure that like is compared with like. The dataset is the largest of its kind and is the first consistent dataset on Olympic costs. Further details on data and methodology are available in Flyvbjerg and Stewart (2012).

*Table 1: Final outturn sports-related cost of the Olympic Games 1960-2016, in 2015 USD; not including road, rail, airport, hotel, and other infrastructure, which often cost more than the Games themselves*

| Games | Country | Events | Athletes | Cost, billion USD |
|---|---|---|---|---|
| **Summer**: | | | | |
| Rome 1960 | Italy | 150 | 5338 | n/a |
| Tokyo 1964 | Japan | 163 | 5152 | 0.282 |
| Mexico City 1968 | Mexico | 172 | 5516 | n/a* |
| Munich 1972 | Germany | 195 | 7234 | 1.009 |
| Montreal 1976 | Canada | 198 | 6048 | 6.093 |
| Moscow 1980 | Soviet Union | 203 | 5179 | 6.331 |
| Los Angeles 1984 | United States | 221 | 6829 | 0.719 |
| Seoul 1988 | South Korea | 237 | 8397 | n/a |
| Barcelona 1992 | Spain | 257 | 9356 | 9.687 |
| Atlanta 1996 | United States | 271 | 10318 | 4.143 |
| Sydney 2000 | Australia | 300 | 10651 | 5.026 |
| Athens 2004 | Greece | 301 | 10625 | 2.942 |
| Beijing 2008 | China | 302 | 10942 | 6.810 |
| London 2012 | United Kingdom | 302 | 10568 | 14.957 |
| Rio 2016 | Brazil | 306 | 10500 | 13.692 |
| ***Average, Summer*** | **-** | **239** | **8177** | **5.974** |
| ***Median, Summer*** | **-** | **237** | **8397** | **5.560** |
| | | | | |
| **Winter**: | | | | |
| Squaw Valley 1960 | United States | 27 | 665 | n/a |
| Innsbruck 1964 | Austria | 34 | 1091 | 0.022 |
| Grenoble 1968 | France | 35 | 1158 | 0.888 |
| Sapporo 1972 | Japan | 35 | 1006 | 0.117 |
| Innsbruck 1976 | Austria | 37 | 1123 | 0.118 |
| Lake Placid 1980 | United States | 38 | 1072 | 0.435 |
| Sarajevo 1984 | Yugoslavia | 39 | 1272 | n/a* |
| Calgary 1988 | Canada | 46 | 1432 | 1.109 |
| Albertville 1992 | France | 57 | 1801 | 1.997 |
| Lillehammer 1994 | Norway | 61 | 1737 | 2.228 |
| Nagano 1998 | Japan | 68 | 2176 | 2.227 |
| Salt Lake City 2002 | United States | 78 | 2399 | 2.520 |
| Torino 2006 | Italy | 84 | 2508 | 4.366 |
| Vancouver 2010 | Canada | 86 | 2566 | 2.540 |
| Sochi 2014 | Russia | 98 | 2780 | 21.890 |
| ***Average, Winter*** | **-** | **55** | **1652** | **3.112** |
| ***Median, Winter*** | **-** | **46** | **1432** | **1.997** |

*) Mexican Peso and Yugoslavian dinar experienced hyperinflation during or after the Games.

n/a: not available.

*Why the Olympics Blow Up © Copyright by the authors, all rights reserved, AAM* 28*Table 2: Sports-related cost per event and per athlete in the Olympics 1960-2016, million 2015 USD*

| Games | Country | Cost per event, mio. USD | Cost per athlete, mio. USD |
|---|---|---|---|
| ***Summer****:* | | | |
| Tokyo 1964 | Japan | 1.7 | 0.1 |
| Munich 1972 | Germany | 5.2 | 0.1 |
| Montreal 1976 | Canada | 30.8 | 1.0 |
| Moscow 1980* | Soviet Union | 31.2 | 1.2 |
| Los Angeles 1984 | United States | 3.3 | 0.1 |
| Barcelona 1992 | Spain | 37.7 | 1.0 |
| Atlanta 1996 | United States | 15.3 | 0.4 |
| Sydney 2000 | Australia | 16.8 | 0.5 |
| Athens 2004 | Greece | 9.8 | 0.3 |
| Beijing 2008 | China | 22.5 | 0.6 |
| London 2012 | United Kingdom | 49.5 | 1.4 |
| Rio 2016† | Brazil | 44.7 | 1.3 |
| ***Average, Summer*** | **-** | ***22.4*** | ***0.7*** |
| ***Median, Summer*** | **-** | ***19.7*** | ***0.6*** |
| | | | |
| ***Winter****:* | | | |
| Innsbruck 1964 | Austria | 0.6 | 0.02 |
| Grenoble 1968 | France | 25.4 | 0.8 |
| Sapporo 1972 | Japan | 3.4 | 0.1 |
| Innsbruck 1976 | Austria | 3.2 | 0.1 |
| Lake Placid 1980 | United States | 11.5 | 0.4 |
| Calgary 1988 | Canada | 24.1 | 0.8 |
| Albertville 1992 | France | 35.0 | 1.1 |
| Lillehammer 1994 | Norway | 36.5 | 1.3 |
| Nagano 1998 | Japan | 32.7 | 1.0 |
| Salt Lake City 2002 | United States | 32.3 | 1.1 |
| Torino 2006 | Italy | 52.0 | 1.7 |
| Vancouver 2010 | Canada | 29.5 | 1.0 |
| Sochi 2014 | Russia | 223.4 | 7.9 |
| ***Average, Winter*** | **-** | ***39.2*** | ***1.3*** |
| ***Median, Winter*** | **-** | ***29.5*** | ***1.0*** |

*) The Moscow 1980 Summer Games were boycotted by 65 nations as a protest against the 1979 Soviet invasion of Afghanistan. The number of participating athletes was therefore lower than anticipated, driving up cost per athlete.

†) Final updates for Rio 2016 costs have been used, according to the 6th and final update of the responsibility matrix, published on June 14, 2007. The matrix has been criticized in the press for omitting some sports related cost and thus the true cost are likely higher.

*Table 3: Sports-related cost overrun, Olympics 1960-2016; calculated in local currencies, real terms*

| Games | Country | Cost overrun % |
|---|---|---|
| ***Summer:*** | | |
| Montreal 1976 | Canada | 720 |
| Barcelona 1992 | Spain | 266 |
| Atlanta 1996 | United States | 151 |
| Sydney 2000 | Australia | 90 |
| Athens 2004 | Greece | 49 |
| Beijing 2008 | China | 2 |
| London 2012 | United Kingdom | 76 |
| Rio 2016 | Brazil | 352 |
| ***Average, Summer*** | - | ***213*** |
| ***Median, Summer*** | - | ***120*** |
| | | |
| ***Winter:*** | | |
| Grenoble 1968 | France | 181 |
| Lake Placid 1980 | United States | 324 |
| Sarajevo 1984 | Yugoslavia | 118 |
| Calgary 1988 | Canada | 65 |
| Albertville 1992 | France | 137 |
| Lillehammer 1994 | Norway | 277 |
| Nagano 1998 | Japan | 56 |
| Salt Lake City 2002 | United States | 24 |
| Torino 2006 | Italy | 80 |
| Vancouver 2010 | Canada | 13 |
| Sochi 2014 | Russia | 289 |
| ***Average, Winter*** | - | ***142*** |
| ***Median, Winter*** | - | ***118*** |

*Table 4: The Olympic Games have the largest cost overrun of any type of large-scale project, real terms*

|  | **Roads** | **Bridges, tunnels** | **Energy** | **Rail** | **Dams** | **IT** | **Olympics** |
|---|---|---|---|---|---|---|---|
| **Cost overrun** | 20% | 34% | 36% | 45% | 90% | 107% | 172% |
| **Frequency of cost overrun** | 9 of 10 | 9 of 10 | 6 of 10 | 9 of 10 | 7 of 10 | 5 of 10 | 10 of 10 |
| **Schedule overrun** | 38% | 23% | 38% | 45% | 44% | 37% | 0% |
| **Schedule length, years** | 5.5 | 8.0 | 5.3 | 7.8 | 8.2 | 3.3 | 7.0 |

Table 5: Events that follow power-law distributions and the law of regression to the tail, with alpha-values. The lower the alpha-value, the fatter the tail and the higher the risk for extreme values

| Type of event | Alpha |
|---|---|
| Earthquakes (intensity as Richter Scale maximum peak, Clauset et al. 2006) | 0.6 |
| Cybercrime (financial loss, Maillart and Sornette 2010) | 0.6 |
| Wars (number of battle deaths per capita, Newman 2005) | 0.7 |
| IT procurement (percentage cost overrun, Flyvbjerg et al. in progress) | 0.9 |
| Bankruptcies (percentage of firms per year per industry, Hong et al. 2007) | 1.1 |
| Floods (volume of water, Malamud and Turcotte, 2006) | 1.1 |
| Forest fires (size of area affected, Clauset et al. 2009) | 1.2 |
| *Olympic Games (percentage cost overrun, present study)* | *1.2-1.7* |
| Terrorist attacks (number of deaths, Clauset et al. 2006) | 1.4 |

*Table 6: Sample mean and true mean for Olympics cost overrun*

| **Estimation** | **Mean** |
|---|---|
| Sample (raw) | 2.72 |
| Lognormal | 2.69 |
| MLE Pareto 1 | 8.46 |
| Alt Pareto 2 | 3.59 |

*Figure 1: Average GDP growth for host nations before, at, and after the Olympic Games, for all Games since 1960. Year 0 = year of the Games (moving average, Loess regression).*

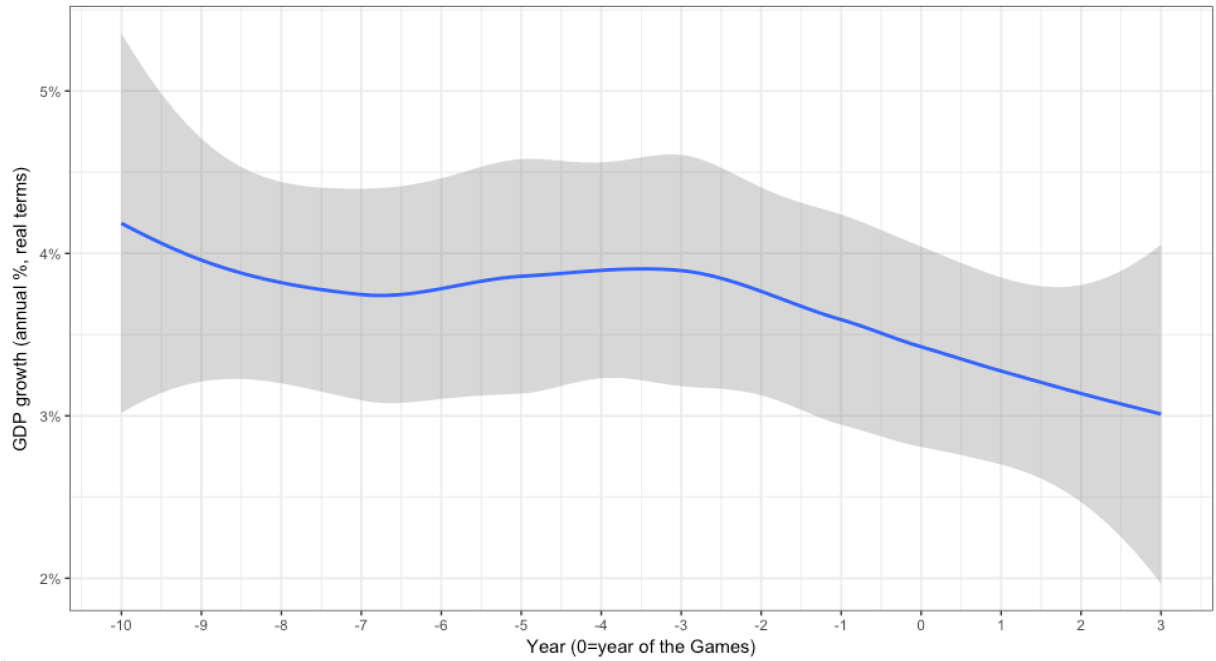

*Figure 2: Time series of outturn sports-related cost for Olympics 1960-2016*

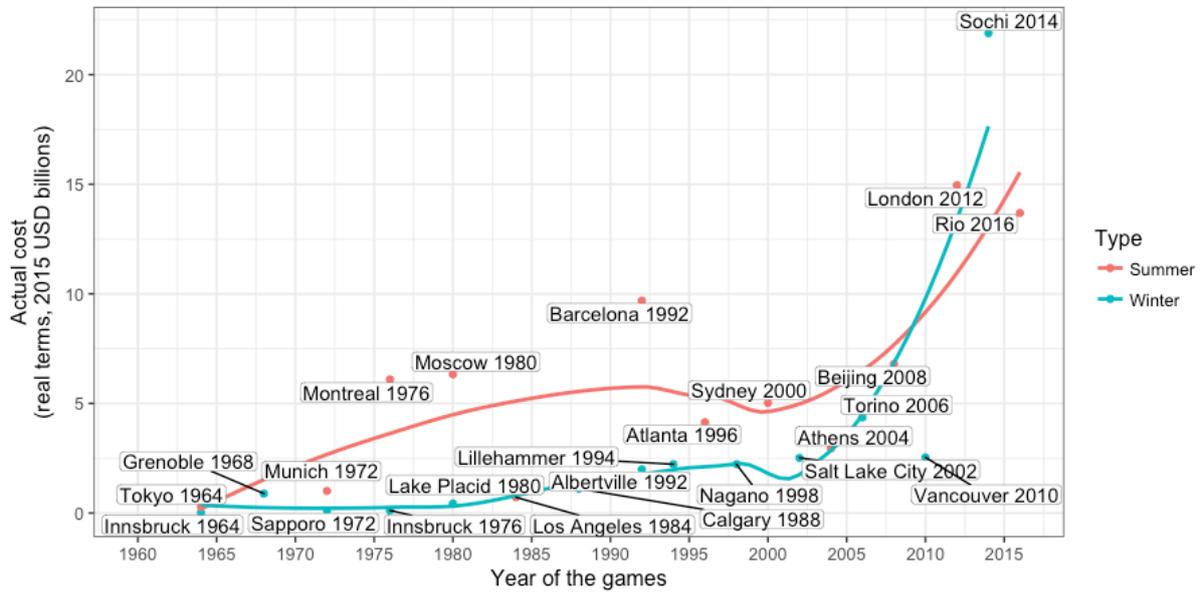

*Figure 3: Time series of sports-related cost per athlete for Olympics 1960-2016, with and without Sochi 2014 as outlier*

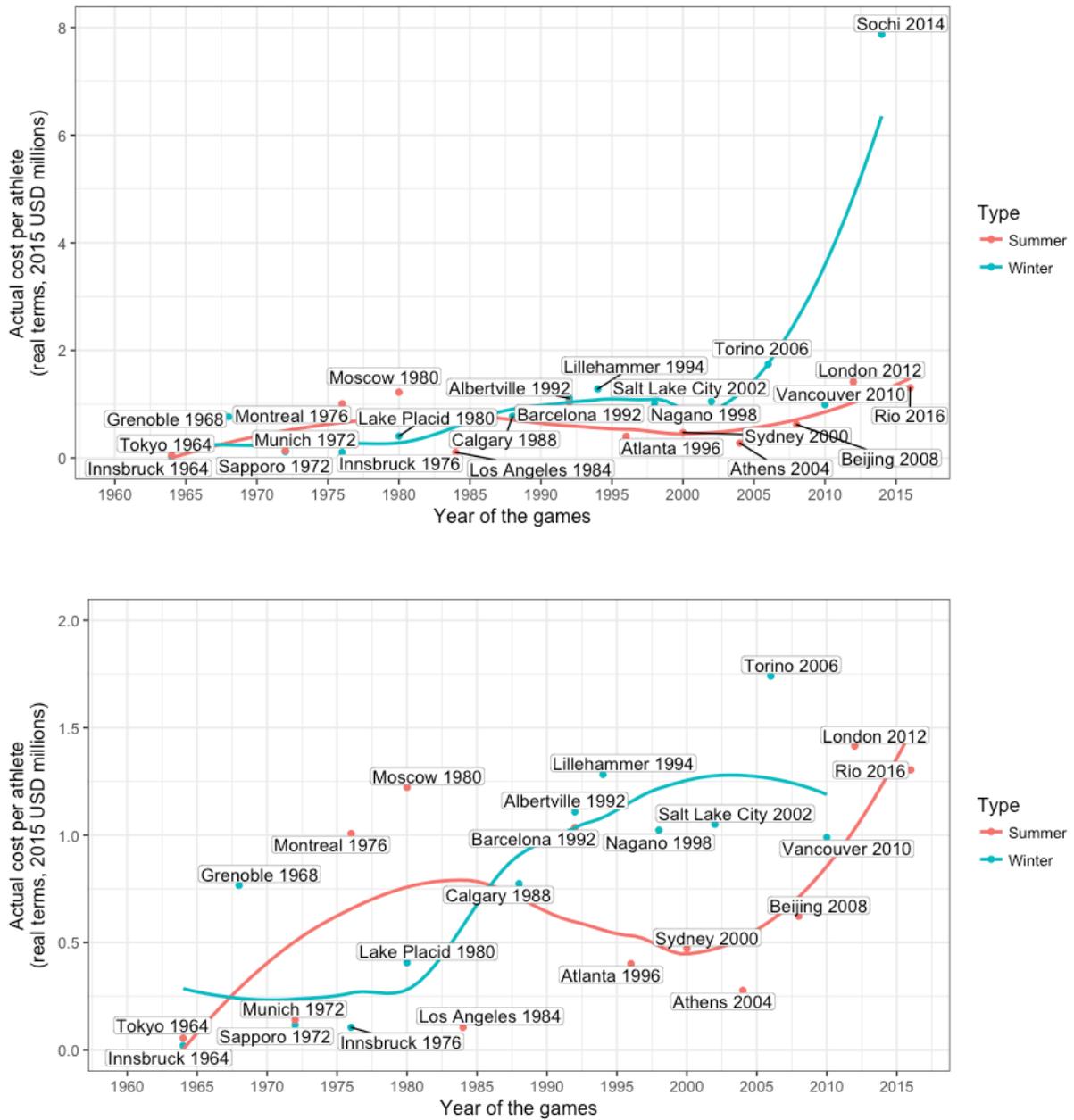

*Figure 4: Log-Log Plot of the data P>x , allowing the visual fitting of a variety of distributions, all fat-tailed*

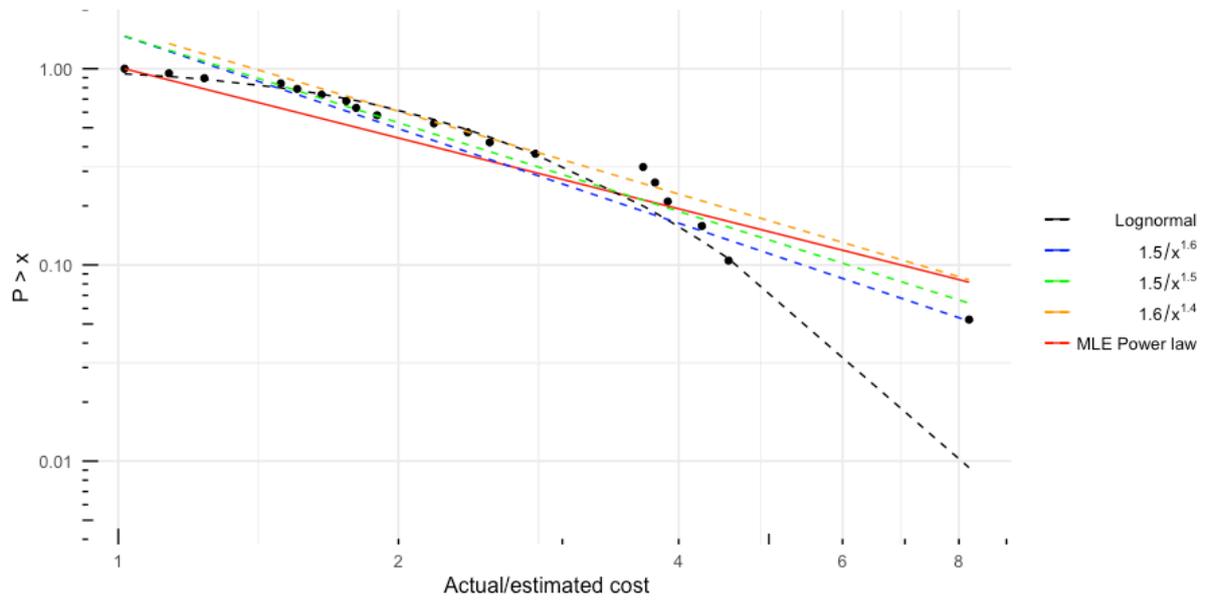

Source: Authors' data.

*Figure 5: Eight states of randomness, from zero to infinite.. Thin-tailed distributions, with low randomness, are at the bottom of the figure; fat-tailed distributions, with extreme randomness, at the top (based on Taleb 2020: 27)*

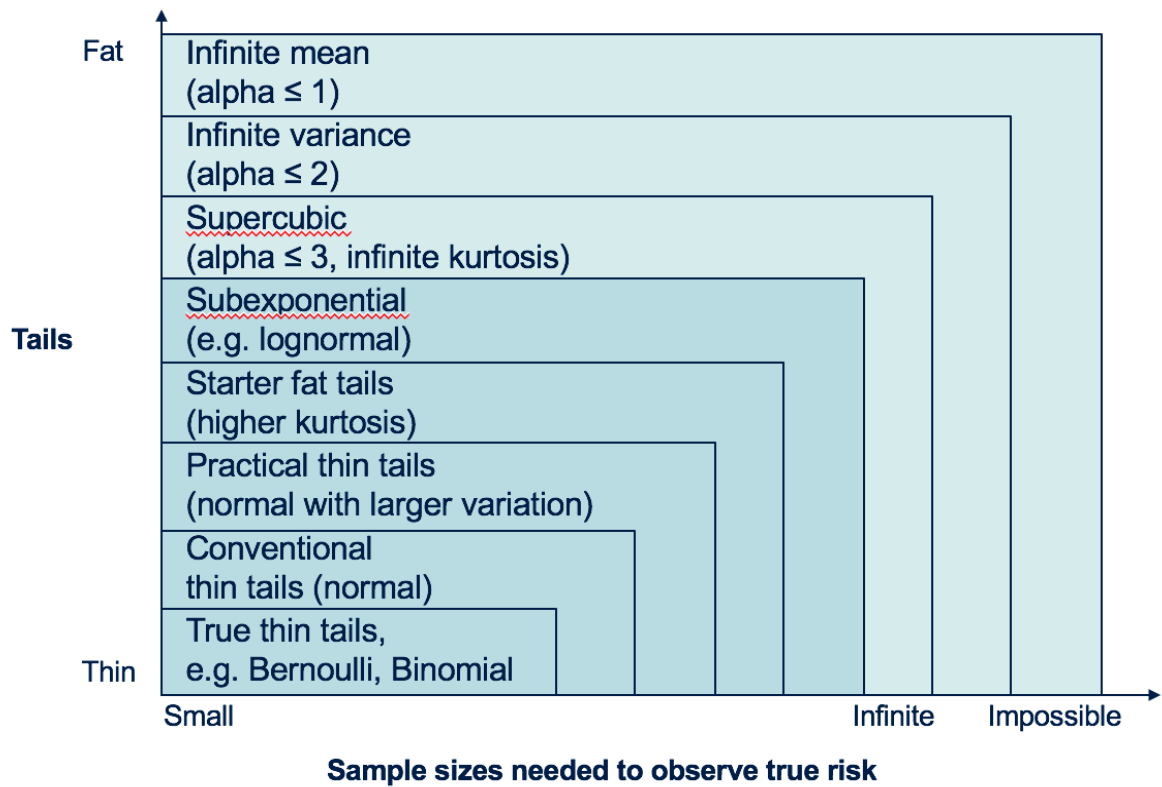

**Endnotes**

[1] All dollar figures are USD at 2015 level, unless otherwise stated.

[2] The Paralympic Games are not included here because they became fully integrated with the Olympic Games relatively recently and therefore do not compare across the period we study.

[3] It should be mentioned that data on final outturn cost for the Rio 2016 Summer Games were published on June 14, 2017 in the 6th and final update of Rio's responsibility matrix (Rio 2016 Organizing Committee and Autoridade Pública Olímpica 2017). Rio's cost accounting has been criticized in the press (Nogueira 2017) for omitting some sports related cost and the true cost might be higher.

[4] Extreme values, like those found here, are often erroneously seen as statistical outliers that are "a nuisance and must be removed" before analysis (Aguinis 2014: 3). Godfrey et al. (2009) is an example of this approach, with more examples presented in Flyvbjerg et al. (2018). But removal of extreme values is nearly always the wrong strategy. It is justified only if such values come from a different population of uncertainty than non-extreme observations, for instance due to faulty measurements or experimenter bias, which is not the case for the present dataset. Blanket removal of extreme values, as is common when extreme values are mistaken for outliers, is bad practice that leads to big mistakes when extreme values are large, as they tend to be (Steele and Huber 2004). Aguinis (2014) and Flyvbjerg et al. (2018) found that the mistakes are compounded by researchers typically being vague or not transparent about how so-called outliers are defined, identified, and removed, undermining the validity of results. Removal of extreme values before analysis is like trying to understand how best to earthquake proof buildings without taking the biggest earthquakes into account—not a good idea. We did not remove any extreme values in the dataset, needless to say. If outliers are defined as abnormal values that should be removed, then there are no outliers in the present dataset. We prefer the term "extreme value" to that of "outlier," due to the latter term's association with misguided outlier removal.

[5] Significance is here defined in the conventional manner, with $p \leq 0.05$ being significant, $p \leq 0.01$ very significant, and $p \leq 0.001$ overwhelmingly significant.

[6] We also did this fit for the Winter and Summer Games separately and found no statistically significant difference between the two, wherefore their data should be, and are, pooled.

[7] The true mean is sometimes also called the "shadow" mean.

[8] Different mechanisms have been argued to generate such amplification and power law distributions, for instance self-organized criticality (Bak et al. 1987), over-optimization (Mandelbrot 1953), and preferential attachment (Gabaix 2009). No fully valid and reliable way exists for how to establish the existence of generative mechanisms for power law distributions. Attempts at determining such mechanisms are therefore likely to contain elements of arbitrariness and speculation.

[9] Due to large cost overruns and worries about environmental impacts, citizens of Colorado called into question the wisdom of hosting the Games. A referendum was held, and voters rejected the Games by a large margin. Innsbruck, Austria, stepped in to replace Denver, hosting the Winter Olympics for a second time. The person who led the campaign against the Denver Games went on to become governor of Colorado (Berg 2016).

[10] In designing and upholding the specs for the Games, the IOC has no incentive to keep costs down but quite the opposite. Cost overrun is picked up by the host. The IOC makes its profit directly from the revenues, so any improvement in specs that leads to improved revenues is a positive for the IOC, irrespective of its cost. This lack of liability for costs for the IOC is deeply unhealthy, because it drives up (amplifies) costs.

[11] The Blank Check Syndrome is a case of preferential attachment, which is a type of behavior where some quantity, e.g., money, is distributed among a number of entities (e.g., individuals or organizations) according to how much they already have, so that those who already have much receive more than those who have less. In project management, the dynamics of preferential attachment are known from escalation of commitment (Sleesman et al. 2012). This is the situation where

managers – even when faced with negative outcomes – keep rationalizing and funding their decisions rather than changing their course of action. This may happen, for instance, when managers think that a point of no return has been reached or that sunk costs are too high to opt out. As a consequence, funds keep being allocated to a project mainly because a lot of money was already allocated to it and no one wants to reverse the original decision. Preferential attachment and escalation of commitment have been shown to lead to power-law outcomes (Yule 1925, Farmer and Geanakoplos 2004, Newman 2005, Gabaix 2009, Sleesman et al. 2012). Spending at the Olympics is as pure a case of preferential attachment and escalation of commitment that one can find, with complete lock in and a point of no return reached more than seven years before delivery.

[12] The more tight the budget for a Games (measured as cost per athlete and cost per event in bid, with lower bid cost indicating a tighter budget) the higher the percentage cost overrun, although the correlation is not statistically significant, due to small sample sizes caused by the innate rarity of the Games.

[13] Before 1976, Games were delivered 1-2 years faster than this.

[14] Several cities have hosted the Games more than once, but over the past century only Innsbruck (1964, 1976) has hosted multiple games within 20 years of each other. This happened under the special circumstance that Denver in 1972 unexpectedly decided to abandon the 1976 Winter Games, as mentioned previously. Innsbruck then offered to step in to save the Games, being able to do so at relatively short notice because they had hosted the Games eight years earlier. Innsbruck already had many of the venues that were needed. Nevertheless, Innsbruck 1976 ended up costing more than five times as much as Innsbruck 1964, in constant terms.